\documentclass[a4paper,11pt]{article}
\usepackage{jheppub}
\usepackage{amsfonts}
\usepackage{amsmath,amssymb,amscd}
\usepackage{appendix,marginnote,tikz,pgf,mathtools}
 \usepackage{comment}

\usepackage[numbers]{natbib}
\def\cM{\mathcal{M}}

\def\ll{\left\langle}
\def\rr{\right\rangle}

\def\bean#1\enan{\begin{align*}#1\end{align*}}
\def\bea#1\ena{\begin{align}#1\end{align}}

\def\be{\begin{equation}}
\def\ee{\end{equation}}
\def\ben{\begin{equation*}}
\def\een{\end{equation*}}

\DeclareMathOperator\acosh{arccosh}
\def\pd{\partial}
\usepackage{mathtools}

\newcommand{\DeclareAutoPairedDelimiter}[3]{%
	\expandafter\DeclarePairedDelimiter\csname Auto\string#1\endcsname{#2}{#3}%
	\begingroup\edef\x{\endgroup
		\noexpand\DeclareRobustCommand{\noexpand#1}{%
			\expandafter\noexpand\csname Auto\string#1\endcsname*}}%
	\x}
\DeclareAutoPairedDelimiter{\p}{(}{)}

\title{Dual  Frobenius manifolds of minimal gravity on disk}

\author{Aditya Bawane,}
\author{Hisayoshi Muraki}
\author{and Chaiho Rim}
\affiliation{Department of Physics, 
Sogang University, Seoul 04107, Korea}

\emailAdd{abawane@sogang.ac.kr}
\emailAdd{hmuraki@sogang.ac.kr}
\emailAdd{rimpine@sogang.ac.kr}

\abstract{
Liouville field theory approach to 2-dimensional  gravity 
possesses the duality ($b \leftrightarrow b^{-1}$).
The matrix counterpart of minimal gravity $\cM(q,p)$ ($q<p$ co-prime) 
is effectively described on $A_{q-1}$ Frobenius manifold, 
which may exhibit a similar duality $p\leftrightarrow q$, 
and allow a description on  $A_{p-1}$ Frobenius manifold. 
We have positive results from the bulk one-point 
and the bulk-boundary two-point correlations  on disk
that the dual description of the Frobenius manifold  works
for the unitary series $\cM(q, q+1)$.
However, for the Lee-Yang series $\cM(2, 2q+1)$ on disk 
the duality is checked only partially. 
The main difficulty lies in the absence of a canonical description 
of trace in the continuum limit. 
}
	
\begin{document} 
\maketitle
\flushbottom
	
\section{Introduction} 

The Liouville field theory shows that there is a 
weak-strong coupling constant duality $b \leftrightarrow b^{-1}$
in correlation functions.
This is demonstrated in DOZZ 3-point correlation on sphere  \cite{Dorn:1994xn,Zamolodchikov:1995aa,Teschner:1995yf}
which is invariant under 
\be\label{eq:bulkduality}
b\rightarrow \tilde{b} = b^{-1}\,,\quad
\mu\rightarrow \tilde{\mu} = \frac{(\pi\mu\gamma(b^2))^{b^{-2}}}{\pi\gamma(b^{-2})} ,
\ee
where $\mu$ is the cosmological constant 
and $\gamma(x) = \Gamma(x)/\Gamma(1-x)$. 
The same duality holds for the one-point correlation 
on disk \cite{FZZ,T} with the boundary parameter $s$ unchanged 
where the boundary cosmological constant  $\mu_B$ is is defined as
\be 
\mu_B = \kappa\cosh\pi b s,
\ee
where $\kappa = \sqrt\frac{\mu}{\sin \pi b^2}$.
The minimal Liouville gravity  $\cM(q,p)$ with $q<p$ co-prime
is also effectively described by $A_{q-1}$ Frobenius coordinates \cite{BDM}.
The connection with Liouville gravity is given as $b^2=(q/p) <1$.
Therefore, one might naturally wonder if the same Liouville gravity 
is equally described by $A_{p-1}$ Frobenius coordinates. 
One may find a brief description of this duality property of matrix model in \cite{Ginsparg:1990zc}
and a detailed check for Lee-Yang series on sphere \cite{Belavin:2015ffa,Aleshkin} which shows that careful analysis is needed. 

In this paper we concentrate on testing the duality of the minimal gravity on a disk.
In section 2, we provide the general structure of 2-dimensional gravity 
on a sphere and on a disk using the  coordinates of Frobenius manifold.
In section 3, we check the dual property of unitary models by using bulk one-point correlation and bulk-boundary two-point correlation on a disk. 
We find that flat coordinates of the original and its dual Frobenius manifold provide a convenient path for evaluating correlation numbers on a sphere as well as on a disk. 
In section 4, we check the dual property of Lee-Yang series (non-unitary case) on a disk.
However, the duality is  confirmed only partially.
The difficulty lies in finding the proper contour integration 
which corresponds to summing up the orthonomal states available in the random hermitian matrix model.
Section 5 is the conclusion. 
The appendix contains useful formulae of the Bessel function and the Jacobi polynomials, and
the computation of resonance relations for the original and its dual Frobenius manifolds
obtained from the orthogonality of two-point correlations on a sphere.

\section{Frobenius manifolds in random matrix model of minimal gravity}
\subsection{Frobenius manifold and Liouville correlation numbers}

The matrix degree of freedom of random matrix model is represented by an operator $Q$ in the continuum limit.
It is defined as a differential operator 
\cite{D},
whose Fourier transform reads
\be
	Q(y) = y^q + \sum_{i=1}^{q-1} y^{q-1-i} u_i,
\ee
where some terms irrelevant for our purpose (i.e., calculations on the sphere and disk) are dropped.
Here $y$ stands for the Fourier space coordinate and $u_i$'s form a system of co-ordinates for a 
$(q-1)$-dimensional Frobenius manifold $A_{q-1}$. Details may be found in \cite{BDM,BB} 
and some are summarized in appendix A.
The flat co-ordinates on $A_{q-1}$ are defined by $v_i = \theta_{i,0}$ with
\begin{equation} \label{eq:flatcoord}
	\theta_{i,k} = -\frac{1}{\p{i/q}_{k+1}}\underset{y=\infty}{\mathrm{Res}}Q^{k+\tfrac{i}{q}}(y),
\end{equation}
where $k\in \{0\}\cup \mathbb{N}$ and $\p{a}_n =\Gamma(a+n)/\Gamma(a)$ is 
the Pochhammer symbol.

Minimal Liouville gravity $\cM(q,p)$ is specified by two co-prime integers $p$ and $q$ $(q<p)$, where $q$ is identical to the order of $Q$. $Q$ and its conjugate $P = Q^{p/q}+\dots$ are required to satisfy the Douglas string equation $[P,Q]=1$
which is effectively described in terms of an action principle \cite{Ginsparg:1990zc} where the action is given by
\be\label{eq:Sqp}
	S_{q/p}(v)=\underset{y=\infty}{\text{Res}}
	\left(Q^{(p+q)/q}+\sum_{m, n} t^{m,n}Q^{|pm-qn|/q}\right),
\ee
where $1\leq m <q$ and $1\leq n <p$. Each KdV parameter $t^{m,n}$
has ${\mathbb Z}_2$ symmetry and  is identified with $t^{q-m,p-n}$ and is only defined up to a normalization factor. The number of these parameters is equal to the number of allowable operators in the minimal model. The string equation reduces to the equation of motion,
\be 
\frac{\pd S_{q/p}(v)}{\pd v_i}=0.
\ee
which establishes a relation between $v_i$'s and $t^{m,n}$'s.

The KdV parameters $t^{m,n}$ can be identified with
the coupling constants  $\lambda^{m,n}$ of the Liouville gravity, up to resonance terms in general.
The gravitational scaling dimension (powers of the cosmological constant $\mu$) is assigned to the coupling constants as \cite{KPZ}
\be\label{eq:scalingSqp}
	[t^{m,n}]=[\lambda^{m,n}]=\frac{p+q-|mp-nq|}{2q}.
\ee
 Note that $[\lambda^{1,1}]=1$, and therefore $\lambda^{1,1}$ can be identified with the cosmological constant $\mu$.
The assignment implies that the identification of the coupling constants can involve higher order corrections, so that resonance transformation has non-linear terms \cite{BZ}
\be
	t^{m,n} = \lambda^{m,n} + \sum A_{m_1,n_1,m_2,n_2} \lambda^{m_1,n_1}\lambda^{m_2,n_2}
	+\cdots,
\ee
where  $[\lambda^{m,n}]=[\lambda^{m_1,n_1}]+[\lambda^{m_2,n_2}]$ and  the numerical factors $A_{m_1,n_1,m_2,n_2}$ can be determined by an order-by-order calculation.

The bulk correlation numbers on a sphere are generated from the free energy
\cite{BZ,BDM}
\be \label{eq:bulkgenSqp}
	\mathcal{F_\mathrm{sphere}} = \frac12 \int_0^{v_1*} C^1_{bc} 
\frac{\partial S_{q/p}}{\partial v_{b}}\frac{\partial S_{q/p}}{\partial v_{c}} dv_1,
\ee
where $C_{bc}^{a}=\eta^{ad}C_{dbc}$ is the structure constant 
of the Frobenius manifold in the flat coordinate basis. 
The two-point correlations and their dual descriptions are summarized in appendix C.

The correlation numbers on a disk is simply given 
in terms of the generating function \cite{Ishiki:2010wb}
\be	\label{eq:diskgenSqp}
	\mathcal{F_\mathrm{disk}}
	= -\langle \text{tr} \log\,C(Q)\rangle,	
\ee
where the trace is to be replaced by the integration of a 
continuous variable $x$, which has the highest gravitational scaling dimension. 
$C(Q)$ is a polynomial of the operator $Q$ and can be extended to be matrix valued  depending on the 
insertion of boundary operators. The degree of $C(Q)$ in $Q$ fixes 
the boundary condition.

The simplest boundary condition corresponding to an insertion of the identity operator on the disk boundary is obtained by taking
a first order polynomial $C(Q)$ in $Q$, denoted by $C(Q)=\mu_B-Q$.
Here $\mu_B$ is a parameter corresponding to the boundary cosmological constant.
In this case, the generating function reads
\be \label{GeneFunct1}
	\mathcal{F}_{(1)}
	= -\left\langle \text{tr} \log(\mu_B - Q)\right\rangle,
\ee
where we introduce a subscript to indicate the order of $C(Q)$
and the trace is to be replaced by the contour integration over $x$ with 
the highest gravitational scaling dimension.
The variable $x$ corresponds to the continuum limit of the orthonormal states available in the
hermitian matrix model.
A derivative of the generating function with respect to a coupling constant $ \lambda^{m,n}$ yields
a bulk one-point correlation number with $(1,1)$ boundary condition:
\be \label{CorreNumDef}
	\frac{\pd \mathcal{F}_{(1)}}{\pd \lambda^{m,n}}\bigg|_*=\ll O_{m,n}\rr^{(1,1)}(s).	
\ee
The subscript $*$ denotes the on-shell values which 
turn off all coupling constants except $\lambda^{1,1}$ 
and use the Frobenius coordinates as the solution of the Douglas 
string equation. 

The correlation number \eqref{CorreNumDef} has an equivalent but alternative expression
in terms of a function $\mathcal{W}(l)$ introduced in the following. 
A derivative of $\mathcal{F}_{(1)}$ with respect to $\mu_B$ gives
\be \label{DefofOmegaGeneral}
	\omega(\mu_B)
	\equiv\frac{\pd \mathcal{F}_{(1)}}{\pd \mu_B}=-\ll\text{tr}\frac{1}{\mu_B-Q} \rr
	\equiv- \int_0^\infty dl \ e^{-l\mu_B} \mathcal{W}(l),	
\ee
where $\mathcal{W}(l)$ is defined by
\be	\label{DefofWGeneral}
	\mathcal{W}(l)=\ll \text{tr} \ e^{lQ} \rr= \int_{x_1}^\infty dx \int_{i\mathbb{R}} dy \ e^{lQ}.
\ee
Here $x_1$ is identified with the coupling constant $\lambda^{m_1,n_1}$ of the highest scaling dimension.
By definition, $\omega(\mu_B)$ is the Laplace transform of $\mathcal{W}(l)$ 
where its integration contour of $l$ from 0 to infinity is 
to be properly defined so that the Laplace transformation is convergent.  
Using this object $\mathcal{W}(l)$, an alternative expression for the correlation number is obtained by
\be \label{OnePointLaplace}
	\ll O_{m,n}\rr^{(1,1)}(s)
	= \int_0^\infty \frac{dl}{l} \ e^{-l\mu_B} \ll O_{m,n}\rr^{(1,1)}_l,
\ee
where
\be \label{CorrNumbInL}
	\ll O_{m,n}\rr^{(1,1)}_l =\frac{\pd \mathcal{W}(l)}{\pd \lambda^{m,n}}\bigg|_*.
\ee
Therefore $\mathcal{W}(l)$ plays the role of generating function with a given boundary length $l$.
Thus we may refer to $\mathcal{W}(l)$ as generating function, as long as no confusion arises.
Accordingly, we will also call the objects like \eqref{CorrNumbInL} correlation numbers hereafter.

The bulk-boundary two-point correlation with $(1,1)$ boundary condition
is obtained by taking the derivative of $\mathcal{F}_{(1)}$
with respective to $\mu_B$ in addition to $\lambda^{m,n}$:  
\be
\left.
	\frac{\pd}{\pd \mu_B} \frac{\pd \mathcal{F}_{(1)}}{\pd \lambda^{m,n}} \right|_* 
	 = \ll B O_{m,n}  \rr^{(1,1)}, \nonumber
\ee
where $B$ is identified with the boundary operator $B_{1,1}$.

Other boundary conditions are obtained when $C(Q)$ is properly defined.
For example, let us take a second order polynomial in $Q$: 
$C(Q)=Q^2+c_1Q+c_2=(Q-x_{1})(Q-x_{-1})$ 
with $x_{1}+x_{-1}=-c_1$ and $x_1x_{-1}=c_2$,
whose derivative with respect to $c_2$ gives
\be \label{DefofDerOmega2}
	\frac{\pd \mathcal{F}_{(2)} }{\pd c_2} = \frac{\omega(x_{1}) - \omega(x_{-1})}{x_{1}-x_{-1}}
	=- \int_0^\infty dl \  \frac{e^{-lx_{1}}-e^{-lx_{-1}}}{x_{1}-x_{-1}}\mathcal{W}(l)
\ee
where we have expanded $1/C(Q)$ in partial fractions and have used $\omega(x_{\pm1})=-\ll \text{tr} (x_{\pm1}-Q)^{-1} \rr$.
Its on-shell value, which is to be identified with the one-point correlation number for the boundary operator $B_{1,2}$, 
is required to vanish. This imposes a condition \cite{Ishiki:2010wb} on the boundary parameters 
$x_{\pm1} = \kappa \cosh\pi b s_{\pm1}$:
\be
s_{\pm1} = s \pm ib \quad {\rm with} \quad s\in\mathbb{R}.
\ee
The on-shell value of the derivative of $\mathcal{F}_{(2)}$ with respect to $\lambda^{m,n}$ reads
\be
	\frac{\pd \mathcal{F}_{(2)} }{\pd \lambda^{m,n}} \bigg|_*
	= \sum_{i=\pm1}\int_0^\infty \frac{dl}{l}  e^{-lx_{i}} \ll O_{m,n}\rr^{(1,1)}_l 
	= \ll O_{m,n}\rr^{(1,1)}(s_{+1})+\ll O_{m,n}\rr^{(1,1)}(s_{-1}),
\ee
which is identified with the bulk one-point correlation 
with $(1,2)$ boundary  $\ll O_{m,n}\rr^{(1,2)}(s)$. 
One may extend the process to the $(1,k)$ boundary
\be	\label{eq:BC:1k}
	\ll O_{m,n}\rr^{(1,k)}(s) =\sum_{\alpha=-(k-1):2}^{k-1}  \ll O_{m,n}\rr^{(1,1)}(s_\alpha) ,
\ee
where $s_\alpha= s + i \alpha b$ and summation is done in steps of 2.
This result is interpreted as the inner product 
between the boundary state $| (1,k);s \rangle$ 
and the bulk state $|(m,n) \rangle $:
\be
\ll O_{m,n}\rr^{(1,k)}(s)=  \langle  (1,k);s |(m,n) \rangle ,
\ee
and $| (1,k);s  \rangle $  is the superposition of 
trivial boundary state with complex valued $s_\alpha$ \cite{SS}
\be
 | (1,k);s  \rangle  = \sum_{\alpha=-(k-1):2}^{k-1} | (1,1);s_\alpha  \rangle .
\ee

One may compute the bulk-boundary two-point correlation for $(1,2)$ boundary condition as:
\be \label{BulkBoundCorr}
\ll B
O_{m,n} \rr^{(1,2)}(s)
=\frac{\pd}{\pd \lambda^{m,n}}\frac{\pd \mathcal{F}_{(2)} }{\pd c_2}\bigg|_*
=- \int_0^\infty dl \  \frac{e^{-lx_{1}}-e^{-lx_{-1}}}{x_{1}-x_{-1}}\ll O_{m,n}\rr^{(1,1)}_l.	
\ee
An extension to the $(1,k)$ boundary condition is straightforward 
when $C(Q)= \sum_{j=0}^k  c_j Q^{k-j} =\prod_{\alpha=-(k-1):2}^{k-1}(Q-x_\alpha)$ (with $c_0=1$)  is used:
\be \label{eq:BBC:1k}
 \ll BO_{m,n} \rr^{(1,k)}(s) 
=\frac{\pd}{\pd \lambda^{m,n}}\frac{\pd \mathcal{F}_{(k)} }{\pd c_k}\bigg|_*
=- \sum_{\alpha=-(k-1):2}^{k-1} \xi^{(k)}_\alpha (x) \int_0^\infty dl \  e^{-lx_\alpha}  \ll O_{m,n}\rr^{(1,1)}_l,
\ee
where $\xi^{(k)}_\alpha$'s are rational functions of $x_\alpha=\kappa\cosh(\pi b s_\alpha)$ given by
\be
	\xi^{(k)}_\alpha (x)=\prod_{\substack{ \beta=-(k-1):2 \\ \beta\neq\alpha}}^{k-1}(x_\alpha-x_\beta)^{-1}.
\ee 
Note that the one-point correlation ($k\neq1$) 
\be	\label{eq:BdyC:1k}
	\ll B
	 \rr^{(1,k)}(s)=\frac{\pd \mathcal{F}_{(k)} }{\pd c_k} \bigg|_*
	=- \sum_{\alpha=-(k-1):2}^{k-1} \xi^{(k)}_\alpha (x) \omega(x_\alpha),
\ee
vanishes because $s_\alpha=s+i\alpha b$.

The results of bulk-boundary two-point correlations \eqref{eq:BBC:1k}
can be compared with their counterparts of field theoretic approach. 
The point of our interest is whether matrix model approach reproduces their $s$-dependent part (up to normalization) \cite{Hosomichi:2001xc,Bourgine:2011cm}
\be \label{BulkBdry}
	R(P_{m,n},kb,s)
	=-\frac{2^{k-1}\kappa^{k-1} }{\sinh(2\pi P_{m,n}/b)} \prod_{\gamma=1}^{k-1} \sinh^2(i\pi b^2\gamma) 
	\sum_{\alpha=-(k-1):2}^{k-1}\xi^{(k)}_\alpha(x)
	\frac{\sinh (2\pi P_{m,n} s_\alpha)}{\sinh (\pi b s_\alpha)},
\ee
with $P_{m,n}=\frac{b}{2}\frac{|mp-nq|}{q}$ 
and the boundary operator $B$ is identified with $B_{1,k}$.

\subsection{Dual Frobenius manifold}

We can equivalently describe the $\cM(q,p)$ (still $q<p$) minimal gravity on a $A_{p-1}$ Frobenius manifold 
(instead of $A_{q-1}$ Frobenius manifold) with $Q(y) = y^p + \sum_{i=1}^{p-1} y^{p-1-i} u_i$. 
In this case the Douglas action in \eqref{eq:Sqp} is replaced by the dual action
\be\label{eq:Spq}
S_{p/q}(v)=\underset{y=\infty}{\text{Res}}
\left(Q^{(p+q)/p}+\sum_{m, n} t^{m,n}Q^{|pm-qn|/p}\right),
\ee
where, as before, $t^{m,n} = t^{q-m,p-n}$. 
Note that for $q<p$ we use the notation $S_{p/q}$ for the dual action and $S_{q/p}$ for the original action.
The scaling dimension (denoted by curly brackets to distinguish from the $A_{q-1}$ case) of the KdV parameters in this dual action is given by
\be\label{eq:scalingSpq}
\{t^{m,n}\}=\{\lambda ^{m,n}\}=\frac{p+q-|mq-np|}{2p},
\ee
which represents the power of $\tilde\mu$. Note that $\{t^{m,n}\} = \frac{q}{p}[t^{n,m}]$. In particular, $
\{\lambda^{1,1}\} = q/p$ measured in units of $\tilde{\mu}$ and therefore, $\lambda^{1,1}$ 
in this dual description can be identified with the cosmological constant $\mu$.

The bulk generating function on a sphere is similar to \eqref{eq:bulkgenSqp}, with $S_{q/p}$ replaced by $S_{p/q}$.
The duality has been tested for the Lee-Yang series in \cite{Belavin:2015ffa,Aleshkin}. One may use the disk generating function analogous to \eqref{eq:diskgenSqp} with $Q$ of order $q$ replaced by that of order $p$.

The resolvents in \eqref{DefofOmegaGeneral} in both descriptions should be equal, 
but it is emphasized that the inverse Laplace transforms $\mathcal{W}(l)$ will be different from each other.
This is because the boundary coupling constant $\mu_B\propto \sqrt{\mu}\cosh(\pi s b)$ 
is replaced by the dual boundary coupling constant $\tilde\mu_B \propto \sqrt{\tilde\mu}\cosh(\pi s \tilde b)$
with $\tilde b=b^{-1}$. 

The bulk one-point correlations on disk with $(1,1)$ boundary
in \eqref{OnePointLaplace} and \eqref{CorrNumbInL} are expected to behave similarly. 
To be precise, one finds
\be	\label{eq:BulkCorrAq}
	\langle O_{m,n}^{(q-1)}\rangle^{(1,1)}_l  \propto {\kappa}^{\frac{|mp-nq|}{q}}K_{\frac{|mp-nq|}{q}}(\kappa l),
\ee
where $K_\nu(z)$ is the Bessel function. Here we use the notation $\langle O_{m,n}^{(q-1)}\rangle$ to stand for the expectation value of the $(m,n)$ operator using the original $(A_{q-1})$ action.
In the dual $(A_{p-1})$ description we find  
\be	\label{eq:BulkCorrAp}
	\langle O_{n,m}^{(p-1)} \rangle^{(1,1)}_l  
	\propto {\tilde\kappa}^{\frac{|mp-nq|}{p}}K_{\frac{|mp-nq|}{p}}({\tilde\kappa}l).
\ee
Both should result in the same FZZ one-point correlation:
\be	\label{eq:FZZBC}
	\langle O_{m,n}^{(q-1)} \rangle^{(1,1)} (s)
	= \langle O_{n,m}^{(p-1)}\rangle^{(1,1)}(s) 
	\propto {\kappa}^{\frac{|mp-nq|}{q}}\cosh\p{\pi s \frac{|mp-nq|}{\sqrt{pq}}}.
\ee

We can compute dual expectation values  with $(k,1)$ boundary if we use the boundary free energy $ \mathcal{ \tilde F}_{(k)} $ obtained in the dual $A_{p-1}$ description. For example, the bulk one-point correlation corresponding to \eqref{eq:BC:1k} for the case of $(k,1)$ boundary is given by
\be	\label{eq:BC:k1}
	\ll O_{m,n}\rr^{(k,1)}(s) =\sum_{\alpha=-(k-1):2}^{k-1}  \ll O_{m,n}\rr^{(1,1)}(\tilde s_\alpha) ,
\ee
with  $\tilde s_\alpha= s + i\alpha\tilde b$, so that 
\be
	\ll O_{m,n}\rr^{(k,1)}(s)=  \langle  (k,1);s |(m,n) \rangle ,
\ee
and  
\be
	| (k,1);s  \rangle  = \sum_{\alpha=-(k-1):2}^{k-1} | (1,1); \tilde s_\alpha  \rangle .
\ee
In addition, as a counterpart of \eqref{eq:BBC:1k},
the bulk-boundary two point correlation with $(k,1)$ boundary condition
is expected to have the form
\be \label{eq:BBC:k1}
	\frac{\pd}{\pd \lambda^{m,n}}\frac{\pd  \mathcal{\tilde F}_{(k)} }{\pd \tilde{c}_k}\bigg|_*
	= \ll B O_{m,n} \rr^{(k,1)}(s) .
\ee
The boundary one-point correlations $\ll B\rr^{(k,1)}$ have a form analogous to  \eqref{eq:BdyC:1k} and therefore
vanish when $k\neq1$.

\section{Unitary models: $\mathcal{M}(q,q+1)$}
\subsection{$\mathcal{M}(3,4)$} 
\subsubsection{Frobenius manifold $A_2$}

We recapitulate the calculations presented in \cite{Aleshkin:2017yty}, denoting $Q_3$ as $Q$ of degree 3:
\begin{equation}
Q_3(y) = y^3 +  u_1y + u_2.
\end{equation}
The flat co-ordinates are $v_i = \theta_{i,0} = u_i$ $(i=1,2)$ using \eqref{eq:flatcoord}.
The Douglas action is given by a linear combination of 
$\theta_{1,0},\theta_{2,0}, \theta_{2,1}$ and $\theta_{1,2}$:
\begin{equation}
S_{3/4} = t^{1,1}v_1+t^{1,2}v_2 + t^{1,3}\left(\frac{-v_1^3}{18} + \frac{v_2^2}{2}\right) -\left(\frac{-v_1^4}{36} + \frac{v_1v_2^2}{2}\right).
\end{equation}
The scaling dimensions of the KdV parameters are
\begin{equation*}
[t^{1,1}] = 1,\quad [t^{1,2}] = \frac{5}{6}, \quad[t^{1,3}] = \frac{2}{6}.
\end{equation*}
We do not need the correction terms in the resonance transformation since 
$t^{1,n} = \lambda^{1,n} + \mathcal{O}(\lambda^3)$.
The derivatives of $S_{3/4}$ are given by
\begin{equation}
\begin{aligned}
\frac{\partial S_{3/4}}{\partial v_1} &=t^{1,1}-x \quad \textrm{where} \quad x \equiv t^{1,3}\frac{v_1^2}{6}-\frac{v_1^3}{9}+ \frac{v_2^2}{2},\\
\frac{\partial S_{3/4}}{\partial v_2} &=t^{1,2}+ t^{1,3}v_2 - v_1v_2.
\end{aligned}
\end{equation}
The Douglas equations are defined by setting $(\partial S_{3/4}/\partial v_i) = 0$. We choose the following solution to define the on-shell conditions:
\begin{equation}
-(v_{1*})^3 \propto \mu,\hspace{0.5cm}v_{2*} = 0.
\end{equation}

The generating function for correlation numbers on a disk is given by
\begin{equation}
	\mathcal{W}(l)=  \int_{\lambda^{1,1}}^{\infty}dx \int_{i\mathbb{R}}dye^{lQ_3(y,v(x))}.
\end{equation}
The one-point correlation  $\langle {O}_{1,n}\rangle^{(1,1)}_l$ on a disk
 can be obtained by differentiating $\mathcal{W}(l)$ with respect to $\lambda^{1,n}$ and evaluating the result on-shell. For example, see the formulae in appendices A and B,
\begin{equation*}
	\langle {O}_{1,1}\rangle^{(1,1)}_l \propto \int_{i\mathbb{R}}dye^{lQ_3(y,v(x))}|_* 
	\propto \int_{\mathbb{R}}dye^{-il(y^3 - v_*y)}\propto\kappa^{1/3}K_{1/3}(\kappa l),
\end{equation*}
where $\kappa\propto \sqrt{\mu}$. 
This gives the desired $\mu$ and $s$ dependence  as given in \eqref{eq:BulkCorrAq}.
Strictly speaking, the argument of the Bessel function 
differs by a numerical factor. However it can be eliminated by an appropriate rescaling of $l$.

The correlator $\langle {O}_{1,2}\rangle_l^{(1,1)}$ is slightly more involved,
since we need to perform the $x$ integral including an additional factor
$(\partial Q_3/\partial\lambda^{1,2})$ which is obtained by a variation of 
the Douglas equations with respect to $\lambda^{1,2}$. That reads
\begin{equation*}
\begin{aligned}
	\langle {O}_{1,2}\rangle^{(1,1)}_l 
	&\propto \int_{\lambda^{1,1}}^{\infty}dx\int_{i\mathbb{R}}dy\left(l\frac{\partial Q_3}{\partial\lambda^{1,2}}e^{lQ_3(y,v(x))}\right)_* \\
	&\propto \int_{\lambda^{1,1}}^{\infty}dx\int_{i\mathbb{R}}dy\left(\frac{l}{v_1}e^{lQ_3(y,v(x))}\right)_*\\
	&\propto \kappa^{2/3}K_{2/3}(\kappa l).
\end{aligned}
\end{equation*}
Similarly,
\begin{equation*}
\begin{aligned}
	\langle {O}_{1,3}\rangle^{(1,1)}_l 
	&\propto \kappa^{5/3}K_{5/3}(\kappa l).
\end{aligned}
\end{equation*}

Making use of Laplace transformation \eqref{OnePointLaplace}, we see that the bulk-one point correlation numbers
with $(1,1)$ boundary condition reads
\be
\ll O_{1,n}\rr^{(1,1)}(s) \propto {\kappa}^{\frac{|4-3n|}{3}} \cosh\left(\pi b s \frac{|4-3n|}{3}\right),
\ee
for $n=1,2,3$, which reproduces the known result \eqref{eq:FZZBC}.

To find the correlation numbers with $(1,k)$ boundary condition we need the resolvent $\omega(x_\alpha)$ where $x_\alpha=\kappa \cosh(\pi b s_\alpha)$:
\begin{equation*}
\omega(x_\alpha)_* = -\int_{0}^{\infty}dl\,e^{-lx_\alpha}\mathcal{W}(l)_* \propto \kappa^{4/3}\cosh(\pi s_\alpha b^{-1}).
\end{equation*}
The one-point boundary correlation numbers $\ll B\rr^{(1,k)}$ $(k\geq2)$ in terms of $\omega(x_\alpha)$,
given by \eqref{eq:BdyC:1k}, vanish on-shell since we set $s_\alpha=s+i\alpha b$ 
($s\in\mathbb{R}$ and $\alpha\in\mathbb{Z}$).
  The bulk-one point correlation numbers with $(1,k)$ boundary, i.e. $\ll O_{1,n}\rr^{(1,k)}(s)$, 
reduces to a sum of $\ll O_{1,n}\rr^{(1,1)}(s_\alpha)$ as observed in \eqref{eq:BC:1k}.
Finally, we may check the $s$-dependence of the bulk-boundary two-point correlation \eqref{eq:BBC:1k} as follows:
\bea \label{eq:BBC1ks}
 \ll B
 O_{1,n} \rr^{(1,k)}(s) 
&=- \sum_{\alpha=-(k-1):2}^{k-1} \xi^{(k)}_\alpha (x) \int_0^\infty dl \  e^{-lx_\alpha}  \ll O_{1,n}\rr_l
\nonumber\\
&\propto {\kappa}^{\frac{|4-3n|}{3}-1} \sum_{\alpha=-(k-1):2}^{k-1} \xi^{(k)}_\alpha (x)
	\frac{\sinh\left(\pi b s_\alpha \frac{|4-3n|}{3}\right)}{\sinh\left(\pi b s_\alpha\right)},
\ena
which is proportional to $R(P_{1,n},kb,s)$ introduced in \eqref{BulkBdry}, where $P_{1,n}=\frac{b}{2}\frac{|4-3n|}{3}$ and $B$ is identified with $B_{1,k}$.

\subsubsection{Dual Frobenius manifold $A_3$} 

Let us now instead take $Q$ to be of degree 4:
\begin{equation}
	Q_4(y) = y^4 + y^2 u_1 + y u_2 + u_3.
\end{equation}
We find that the flat co-ordinates $v_i$ in this case are related with $u_i$'s by
\begin{equation*}
	v_1 = u_1, \quad v_2=u_2\quad v_3=u_3-\frac{u_1^2}{8}.
\end{equation*}
The Douglas action is given by a linear combination of $\theta_{1,0},\theta_{2,0},\theta_{1,1}$ and $\theta_{3,1}$:
\be
	S_{4/3} 
	= t^{1,1}v_1+t^{2,1}v_2 + t^{1,3}\p{v_1v_3 + \frac{v_2^2}{2}}-\p{\frac{v_1^4}{192} 
	+ \frac{v_1v_2^2}{2}+ \frac{v_3^2}{2}}.
\ee
The dual scaling dimensions of the KdV parameters are
\begin{equation*}
\{t^{1,1} \} = \frac{3}{4},\quad \{t^{1,2} \}  = \frac{5}{8}, \quad\{t^{1,3} \}  = \frac{2}{8}.
\end{equation*}
Since the ratios of the scaling dimensions is the same as in the $A_2$ description, 
there continues to be no non-trivial resonance transformation up to the order we need.
The derivatives of $S_{4/3}$ are given by
\begin{equation}
\begin{aligned}
\frac{\partial S_{4/3}}{\partial v_1} 
&=t^{1,1}-x\quad \textrm{where}\quad x = - t^{3,1}v_3 -\frac{v_1^3}{48}+ \frac{v_2^2}{8},\\
\frac{\partial S_{4/3}}{\partial v_2} &=t^{2,1}+ t^{3,1}v_2 - \frac{v_1v_2}{4},\\
\frac{\partial S_{4/3}}{\partial v_3} &= t^{3,1}v_1+v_3.
\end{aligned}
\end{equation}
We choose the following solution to define the on-shell conditions:
\begin{equation}
-(v_{1*})^3 \propto \mu,\quad v_{2*} = 0, \quad v_{3*}=0.
\end{equation}

As before, the correlation number $\langle {O}_{n,1}\rangle^{(1,1)}_l$ is obtained by differentiating
the generating function
\be
	\mathcal{W}(l)=\int_{\lambda^{1,1}}^{\infty}dx \int_{i\mathbb{R}} dy\,e^{lQ_4(y,v(x))},
\ee
with respect to $\lambda^{n,1}$:
\begin{equation*}
	\langle {O}_{1,1}\rangle^{(1,1)}_l 
	\propto \tilde\kappa^{1/4}K_{1/4}(\tilde\kappa l),
\end{equation*}
where $\tilde{\kappa}\propto (v_{1*})^{4/2}\propto \mu^{2/3}\propto \kappa^{4/3}$.
Using $(\partial Q_4/\partial \lambda^{3,1})_* = v_{1}$, we also find that
\begin{equation*}
	\langle {O}_{3,1}\rangle^{(1,1)}_l 
	\propto \tilde\kappa^{5/4}K_{5/4}(\tilde\kappa l).
\end{equation*}
These results are consistent with the duality as suggested in \eqref{eq:BulkCorrAq}-\eqref{eq:FZZBC}.
After Laplace transformation \eqref{OnePointLaplace}, we see that the bulk-one point correlation numbers
with $(1,1)$ boundary condition reads
\bea
\ll O_{n,1}\rr^{(1,1)}(s) 
&\propto {\tilde \kappa}^{\frac{|4-3n|}{4}} \cosh\left(\pi \tilde b s \frac{|4-3n|}{4}\right)\nonumber\\
&=\p{\frac{\tilde \kappa^{1/4}}{\kappa^{1/3}}}^{|4-3n|} 
{ \kappa}^{\frac{|4-3n|}{3}} \cosh\left(\pi \tilde b  s \frac{|4-3n|}{3}b^2\right)\nonumber\\
&\propto {\kappa}^{\frac{|4-3n|}{3}} \cosh\left(\pi b s \frac{|4-3n|}{3}\right)
\propto  \ll O_{1,n}\rr^{(1,1)}(s) \textrm{ in $A_2$ description},
\ena
which reproduces the known result \eqref{eq:FZZBC} (except for $n=2$). Note that $(\tilde \kappa^{1/4}/\kappa^{1/3})$ has zero scaling dimension,
since $[\kappa]=[\mu]/2=1/2$ and $\{\tilde \kappa\}=\{\tilde \mu\}/2=1/2$ 
so that $[\tilde\kappa]=(4/3)\times\{\tilde\kappa\}=2/3$.

However, we run into problems when we try to compute $\langle {O}_{2,1}\rangle^{(1,1)}_l$.
We note that $(\partial Q/\partial t^{21})_* = y/v_{1}$, which makes the integrand for the $y$-integral an odd function, incorrectly giving $\langle {O}_{2,1}\rangle^{(1,1)}_l=0$. This is another instance of a problem that has been pointed out in \cite{Aleshkin:2017yty}. This happens in general when the degree of $Q$ is even and the $m$ in 
$\langle {O}_{m,n}\rangle^{(1,1)}_l$ is even. We will continue to see instances of this in the general case that we deal with in the subsequent sections.

The on-shell value of $\omega(\tilde x_\alpha)$ with  $\tilde x_\alpha=\tilde \kappa \cosh(\pi \tilde b \tilde s_\alpha)$ reads
\begin{equation*}
	\omega(\tilde x_\alpha)_* 
	= -\int_{0}^{\infty}dl\,e^{-l\tilde{x}_\alpha}\mathcal{W}(l)_* 
	\propto \tilde{\kappa}^{3/4} \cosh(\pi \tilde{s}_\alpha \tilde{b}^{-1}).
\end{equation*}
The one-point boundary correlation numbers $\ll B\rr^{(k,1)}$ $(k\geq2)$ vanish because
$\tilde s_\alpha= s + i\alpha\tilde b$. 
The bulk one-point correlation numbers $\ll O_{n,1}\rr^{(k,1)}(s)$ with $(k,1)$ boundary $(k=2,3)$ 
reduces to a sum of $\ll O_{n,1}\rr^{(1,1)}(s_\alpha)$ as in \eqref{eq:BC:k1}.
The bulk-boundary two-point correlation  \eqref{eq:BBC:k1}:
\bea
	\ll B O_{n,1} \rr^{(k,1)}(s) 
	&=- \sum_{\alpha=-(k-1):2}^{k-1} \xi^{(k)}_\alpha (\tilde x) \int_0^\infty dl \  e^{-l\tilde x_\alpha}  
	\ll O_{n,1}\rr_l\nonumber\\
	&\propto {\tilde \kappa}^{\frac{|4-3n|}{4}-1} 
		\sum_{\alpha=-(k-1):2}^{k-1} \xi^{(k)}_\alpha (\tilde x)
		\frac{\sinh\left(\pi \tilde b \tilde s_\alpha 
		\frac{|4-3n|}{4}\right)}{\sinh\left(\pi \tilde b \tilde s_\alpha\right)},	
\ena
is proportional to $R(P_{n,1},k\tilde{b},s)$ with $P_{n,1}=\frac{\tilde{b}}{2}\frac{|4-3n|}{4}=\frac{b}{2}\frac{|4-3n|}{3}$,
which is the same as $P_{1,n}$ of $A_2$ description as mentioned below \eqref{eq:BBC1ks}.
Here $B=B_{k,1}$ and the matrix model approach reproduces the $s$-dependent part of the bulk-boundary two-point correlation.

\subsection{$\cM(q,q+1)$}
\subsubsection{Frobenius manifold $A_{q-1}$}

The Douglas action for the unitary model $\mathcal{M}(q,q+1)$ is given by
\begin{equation}
S_{q/(q+1)} = -\theta_{1,2}+\sum_{i=1}^{q-1}t^{i,i}v_i+\sum_{i=2}^{q-1}\sum_{j=1}^{i-1}t^{i,j}\theta_{i,i-j},
\end{equation}
where the $\theta$'s are derived from $Q_q(y)$ of order $q$.
As mentioned in the introduction, operators in minimal gravity $\cM(q,p)$ are labelled by pairs of integers $(m,n)$ where $1\leq m<q$ and $1\leq n <p$, up to an identification of indices $(m,n) \leftrightarrow (q-m,p-n)$.
Then we need to choose a subset of indices that label all operators without any redundancies.
A convenient choice for $(m,n)$ is requiring $m\geq n$, i.e.,
 the entries on and under the diagonal in the matrix below:
\[
\begin{matrix}
\mathbf{(1,1)} & (1,2) & \dots & (1,q) \\
\mathbf{(2,1)} & \mathbf{(2,2)} & \dots & \vdots\\
\vdots & \vdots & \ddots & \vdots\\
\mathbf{(q-1,1)} & \dots &\mathbf{(q-1,q-1)}& (q-1,q)
\end{matrix}
\]
The gravitational scaling associated with those operators are
\begin{equation*}
[t^{m,n}]=[\lambda^{m,n}] = m-n +\frac{m}{q} \equiv \nu_{m,n},
\end{equation*}
so that the resonance transformation reads
\be \label{tlambda}
	t^{m,n} 
	= \lambda^{m,n} + \sum A^{m,n}_{m_1,n_1} \mu^{\nu_{m,n} - \nu_{m_1,n_1}}\lambda^{m_1,n_1} + 
	O(\lambda^2).
\ee
 We define $x$ by
\be
	\frac{\pd S_{q/(q+1)}}{\pd v_1} = t^{1,1}-x  \quad\textrm{where}\quad x
	=\frac{\pd \theta_{1,2}}{\pd v_1} + \textrm{(terms with $t^{i,j}$ besides  $t^{1,1}$)}.
\ee
On-shell, when all couplings besides $\lambda^{1,1}$ vanish, we have
\begin{equation*}
	x = \frac{\pd \theta_{1,2}}{\pd v_1} =  q\p{\frac{-v_1}{q}}^q.
\end{equation*}

In order to compute the bulk one-point correlation on the disk $\langle{O}_{m,n}\rangle^{(1,1)}_l$, 
we need to evaluate $(\pd Q/\pd \lambda^{m,n})$, which can be computed as follows:
\be \label{eq:dQdlambda}
	\frac{\pd Q_q}{\pd \lambda_{m,n}} = \sum_a \frac{\pd v_a}{\pd \lambda_{m,n}}\frac{\pd Q_q}{\pd v_a}.
\ee
Using \eqref{tlambda}, we can expand the Douglas action as a series in $\lambda$:
\begin{equation*}
	S_{q/(q+1)} = S_0 + \sum_{m,n}\lambda^{m,n}S^{m,n} + O(\lambda^2),
\end{equation*}
where
\begin{equation*}
	S^{m,n} = \theta_{m,m-n} + (\textrm{terms containing $\mu$}).
\end{equation*}
By demanding the vanishing of the variation of the string equation with respect to $\lambda$'s, we find that
\begin{equation}\label{eq:dvdlambda}
	\frac{\pd v_a}{\pd \lambda^{m,n}} 
	= -\frac{\pd S^{m,n}}{\pd v_b} \left[\p{\frac{\pd^2 S_0}{\pd v^2}}^{-1}\right]^{a,b},
\end{equation}
where $[(\pd^2 S_0/\pd v^2)^{-1}]^{a,b}$ stands for the inverse of the Hessian matrix $\p{\pd^2 S_0/\pd v_a\pd v_b}$.
In this case,
\begin{equation}\label{eq:d2sdv2_1}
	\left[\p{\frac{\pd^2 S_0}{\pd v^2}}^{-1}\right]^{a,b} = -\left[\p{\frac{\pd^2 \theta_{1,2}}{\pd v^2}}^{-1}\right]^{a,b} 
	= \delta_{a,b}\frac{1}{q}\p{\frac{-v_1}{q}}^{-q+a}.
\end{equation}
Using orthogonality of the two-point correlations on the sphere \cite{Belavin} we find
\begin{equation}\label{eq:dsmndv}
	\frac{S^{m,n}}{\pd v_a}\propto
	\begin{cases}
		\delta_{m,a} {\mu}^\frac{m-n}{2}P_{\frac{m-n}{2}}^{(0,\frac{m}{q}-1)}\p{\tfrac{2x}{\mu}-1} 
		&\textrm{if $(m-n)$ is even,}\\
		\delta_{m,q-a}x^\frac{m}{q} {\mu}^\frac{m-n-1}{2}P_{\frac{m-n-1}{2}}^{(0,\frac{m}{q})}\p{\tfrac{2x}{\mu}-1} 
		&\textrm{if $(m-n)$ is odd,}
	\end{cases}
\end{equation}
where $P^{(\alpha,\beta)}_n(z)$ stands for the Jacobi polynomial.
The derivation of this is summarized in appendix C.1.
Note that the on-shell value of the derivative $(\pd Q_q/\pd v_a)$ reads 
\begin{equation}\label{eq:dQdv}
	\frac{\pd Q_q}{\pd v_a} 
	= \frac{(-i)^{q-a-1}}{q-a}\p{\frac{v_1}{q}}^{\frac{q-a-1}{2}}T'_{q-a}\p{\tfrac{i}{2}\sqrt{\tfrac{q}{v_1}}y},
\end{equation}
where $T_n(z)$ is the Chebyshev polynomial (see appendix A).
Combining \eqref{eq:d2sdv2_1}, \eqref{eq:dsmndv} and \eqref{eq:dQdv}, the equation \eqref{eq:dQdlambda} yields
\begin{equation}
	\frac{\pd Q_q}{\pd \lambda_{m,n}} =
	\begin{cases}
		\frac{1}{v_1}\p{\frac{-v_1}{q}}^{\frac{1-q+m}{2}}T'_{q-m}\p{\tfrac{i}{2}\sqrt{\tfrac{q}{v_1}}y} 
		{\mu}^\frac{m-n}{2}P_{\frac{m-n}{2}}^{(0,\frac{m}{q}-1)}\p{\tfrac{2x}{\mu}-1} 
		&\textrm{if $(m-n)$ is even,}\\
		\frac{1}{v_1}\p{\frac{-v_1}{q}}^{\frac{1-m}{2}} x^\frac{m}{q}T'_{m}\p{\tfrac{i}{2}\sqrt{\tfrac{q}{v_1}}y} 
		{\mu}^\frac{m-n-1}{2}P_{\frac{m-n-1}{2}}^{(0,\frac{m}{q})}\p{\tfrac{2x}{\mu}-1} 
		&\textrm{if $(m-n)$ is odd.}
	\end{cases}
\end{equation}

As an intermediate step to compute $\langle{O}_{m,n}\rangle^{(1,1)}_l$, 
one needs to perform an integration over $y$: 
\begin{equation}\label{yintegral:Aq}
\int_{i\mathbb{R}} dy\,l\frac{\pd Q_{q}}{\pd \lambda_{n,m}}\ e^{l Q_{q}(y,v(x))}.
\end{equation}
In order for the integral to converge, $l$ is
properly chosen. In addition, we note that the integral identically vanishes when both $q$ and $n$ are even, since the integrand is an odd function of $y$ in this case. This was also noted in \cite{Aleshkin:2017yty}. In every other case, after performing the $y$-integration and up to normalization factors, we obtain
\begin{equation}\label{eq:yint_1}
\begin{aligned}
\sqrt{x}^{\frac{m}{q}-1}{\mu}^\frac{m-n}{2}P_{\frac{m-n}{2}}^{(0,\frac{m}{q}-1)} \p{\tfrac{2x}{\mu}-1} K_{1-\frac{m}{q}}(\sqrt{x}l)l\quad
&\textrm{if $(m-n)$ is even,}\\
\sqrt{x}^{\frac{m}{q}}{\mu}^\frac{m-n-1}{2}P_{\frac{m-n-1}{2}}^{(0,\frac{m}{q})} \p{\tfrac{2x}{\mu}-1} K_{\frac{m}{q}}(\sqrt{x}l)l\quad 
&\textrm{if $(m-n)$ is odd.}
\end{aligned}
\end{equation}

The final step to compute the bulk one-point correlation  $\langle{O}_{m,n}\rangle^{(1,1)}_l$ is performing
the $x$-integration of \eqref{eq:yint_1} from $\mu$ to $\infty$. 
This was done case-by-case in \cite{Aleshkin:2017yty}, and it was verified that it yields the expected answer in all those cases. We fill this gap by proving an integration formula \eqref{eq:JacobiBesselInt}, whose proof is summarized in appendix B. Using this integration formula, we obtain
\begin{equation} \label{eq:BulkOnept:OmnL}
\langle{O}_{m,n}\rangle^{(1,1)}_l \propto \kappa^{m-n+\frac{m}{q}}K_{m-n+\frac{m}{q}}(\kappa l).
\end{equation}

The bulk-one point correlation numbers
with $(1,1)$ boundary condition then reads
\be \label{eq:unitaryOmn}
	\ll O_{m,n}\rr^{(1,1)}(s) \propto \kappa^{m-n+\frac{m}{q}} \cosh\left(\pi b s \frac{(q+1)m-nq}{q}\right),
\ee
which reproduces the known result \eqref{eq:FZZBC}.

The on-shell value of $\omega(x_\alpha)$ with $x_\alpha=\kappa \cosh(\pi b s_\alpha)$ is
\begin{equation*}
\omega(x_\alpha)_* 
= -\int_{0}^{\infty}dl\,e^{-lx_\alpha}\int_0^\infty dx \int_{i\mathbb{R}} dy \ e^{lQ_q(y,v(x))}
\propto \kappa^{(q+1)/q}\cosh(\pi s_\alpha b^{-1}).
\end{equation*}
The one-point boundary correlation numbers $\ll B\rr^{(1,k)}$ given by \eqref{eq:BdyC:1k}
vanish for each $k\geq2$, since $s_\alpha=s+i\alpha b$.
The bulk one-point correlation numbers with $(1,k)$ boundary is easy to check, 
since $\ll O_{m,n}\rr^{(1,k)}(s)$ reduces to a sum of $\ll O_{1,n}\rr^{(1,1)}(s_\alpha)$ as in \eqref{eq:BC:1k}. 

Finally we may check that the bulk-boundary two-point correlation  \eqref{eq:BBC:1k} is
\bea
 \ll B O_{m,n} \rr^{(1,k)}(s) 
&\propto \sum_{\alpha=-(k-1):2}^{k-1} \xi_\alpha (x)
	\frac{\sinh\left(\pi b s_\alpha \frac{(q+1)m-nq}{q}\right)}{\sinh\left(\pi b s_\alpha\right)},	
\ena
with $P_{m,n}=\frac{b}{2}\frac{(q+1)m-nq}{q}$ ($m\geq n$) and $B=B_{1,k}$, 
which is proportional to $R(P_{m,n},kb,s)$.

\subsubsection{Dual Frobenius manifold $A_q$}

The dual action for $\cM(q,q+1)$ is
\begin{equation}
	S_{(q+1)/q} 
	= -\theta_{q,1}+\sum_{i=1}^{q-1}{t}^{i,i} v_i+\sum_{i=1}^{q-2}\sum_{j=i+1}^{q-1}{t}^{i,j}\theta_{i,j-i}.
\end{equation}
Note that the $v$'s and $\theta$'s used in $A_q$ description are distinct from those in $A_{q-1}$ description, 
since these are derived from $Q$ of order $(q+1)$, 
while the ones in the previous subsection were derived from $Q$ of order $q$.

The number of operators in the $\cM(q,q+1)$ model in $A_q$ description is the same as in $A_{q-1}$ description. 
We label operators with indices $(n,m)$, where the two indices related as $(n,m)\leftrightarrow(q+1-n,q-m)$ 
denote the same operator.
We chose a set with $m\geq n$, i.e., the entries on and above the diagonal in the matrix below:
\[
\begin{matrix}
	\mathbf{(1,1)} & \mathbf{(1,2)} & \dots & \mathbf{(1,q-1)} \\
	(2,1) & \mathbf{(2,2)} & \dots & \vdots\\
	\vdots & \vdots & \ddots & \mathbf{(q-1,q-1)}\\
	(q,1) & \dots & \dots& (q,q-1)
\end{matrix}
\]
The gravitational scaling dimensions associated with these operators are
\be
	\{t^{n,m}\}=\{\lambda^{n,m}\} = m-n +\frac{n}{q+1},
\ee
where they are understood as the power of $\tilde{\mu}\propto \mu^{\frac{q+1}{q}}$.
The derivative of the action $S_{(q+1)/q}$ with respect to $v_1$ defines $x$ by
\be
	\frac{\pd S_{(q+1)/q}}{\pd v_1} = t^{1,1}-x \quad \textrm{where} \quad
	x=\frac{\pd \theta_{q,1}}{\pd v_1} + (\textrm{terms with $t^{i,j}$ besides  $t^{1,1}$}).
\ee
The on-shell value of $x$ reads
\begin{equation*}
	x_* = \frac{\pd \theta_{q,1}}{\pd v_1} =  -\frac{q+1}{q}\p{\frac{-v_1}{q+1}}^q.
\end{equation*}
It is convenient to define $\tilde{x}$ such that
\begin{equation}
	\tilde{x} = \tilde{\mu}\p{\frac{x}{\mu}}^\frac{q+1}{q}.
\end{equation}
We observe that $\tilde{x} \propto v_1^{q+1}$.

Using the orthogonality of the two-point correlator on a sphere in the dual description 
(derived in appendix C.2), we find that
\begin{equation}\label{eq:dsnmdv}
	\frac{\pd S^{n,m}}{\pd v_a}\propto
	\begin{cases}
		\delta_{n,a} \tilde{\mu}^\frac{m-n}{2}
		P_{\frac{m-n}{2}}^{(0,\frac{n}{q+1}-1)}\p{\tfrac{2\tilde{x}}{\tilde{\mu}}-1} 
		&\textrm{if $(m-n)$ is even,}
		\\
		\delta_{n,q+1-a} \tilde{x}^\frac{n}{q+1} \tilde{\mu}^\frac{m-n-1}{2}
		P_{\frac{m-n-1}{2}}^{(0,\frac{n}{q+1})}\p{\tfrac{2\tilde{x}}{\tilde{\mu}}-1} 
		&\textrm{if $(m-n)$ is odd.}
	\end{cases}
\end{equation}
The analogue of \eqref{eq:d2sdv2_1} in this case is
\begin{equation}\label{eq:d2sdv2_2}
	\left[\p{\frac{\pd^2 S_0}{\pd v^2}}^{-1}\right]^{a,b} 
	= -\left[\p{\frac{\pd^2 \theta_{q,1}}{\pd v^2}}^{-1}\right]^{a,b} 
	= -\delta_{a,b}\p{\frac{-v_1}{q+1}}^{-q+a}.
\end{equation}
Combining \eqref{eq:dsnmdv}, \eqref{eq:d2sdv2_2} and \eqref{eq:dQdv} gives
\be
	\frac{\pd Q_{q+1}}{\pd \lambda_{n,m}}\propto
	\begin{cases}
		v_1^{\frac{n-q}{2}}\tilde{\mu}^\frac{m-n}{2}
		P_{\frac{m-n}{2}}^{(0,\frac{n}{q+1}-1)}\p{\tfrac{2\tilde{x}}{\tilde{\mu}}-1}
		T'_{q+1-n}\p{\tfrac{i}{2}\sqrt{\tfrac{q+1}{v_1}}y}  
		&\textrm{if $(m-n)$ is even,}
		\\
		v_1^{\frac{1-n}{2}}\tilde{x}^\frac{n}{q+1} \tilde{\mu}^\frac{m-n-1}{2}
		P_{\frac{m-n-1}{2}}^{(0,\frac{n}{q+1})}\p{\tfrac{2\tilde{x}}{\tilde{\mu}}-1}
		T'_{n}\p{\tfrac{i}{2}\sqrt{\tfrac{q+1}{v_1}}y}
		&\textrm{if $(m-n)$ is odd.}
	\end{cases}
\ee
Analogous to \eqref{yintegral:Aq}, we integrate $\int dy\,l\frac{\pd Q_{q+1}}{\pd \lambda_{n,m}}\exp(l Q_{q+1})$. 
Now, the only anomalous case is when $q$ is odd and $n$ is even, in which case the integrand is an odd function of $y$ and the integral identically vanishes. In every other case, the result of this integration up to normalization factors is
\be
\begin{aligned}
	v_1\sqrt{\tilde{x}}^{\frac{n}{q+1}-1}\tilde{\mu}^\frac{m-n}{2}
	P_{\frac{m-n}{2}}^{(0,\frac{n}{q+1}-1)} \p{\tfrac{2\tilde{x}}{\tilde{\mu}}-1} 
	K_{1-\frac{n}{q+1}}(\sqrt{\tilde{x}}l)l\quad
	&\textrm{if $(m-n)$ is even,}
	\\
	v_1\sqrt{\tilde{x}}^{\frac{n}{q+1}}\tilde{\mu}^\frac{m-n-1}{2}
	P_{\frac{m-n-1}{2}}^{(0,\frac{n}{q+1})}\p{\tfrac{2\tilde{x}}{\tilde{\mu}}-1}
	K_{\frac{n}{q+1}}(\sqrt{\tilde{x}}l)l\quad
	&\textrm{if $(m-n)$ is odd.}
\end{aligned}
\ee
The earlier remarks about the normalization of $l$ and a minor abuse of notation apply. 
As a final step, we need to integrate this over $x$ as $\int_{\mu}^{\infty}dx\dots$ in order to obtain 
$\langle\mathcal{O}_{n,m}\rangle^{(1,1)}_l$. But note that  
$\int_{\mu}^{\infty}dx \, v_1\dots \propto \int_{\tilde\mu}^{\infty}d\tilde x \dots$. 
After performing this change of integration variable, we may simply use \eqref{eq:JacobiBesselInt} 
and observe that in either case we obtain
\be
	\langle{O}_{n,m}\rangle^{(1,1)}_l
	=\tilde{\kappa}^{m-n+\frac{n}{q+1}}K_{m-n+\frac{n}{q+1}}(\tilde{\kappa}l).
\ee

The bulk-one point correlation numbers with $(1,1)$ boundary condition reads
\be \label{eq:unitaryOnm}
	\ll O_{n,m}\rr^{(1,1)}(s) \propto \tilde\kappa^{m-n+\frac{n}{q+1}} \cosh\left(\pi \tilde b s \frac{m(q+1)-nq}{q+1}\right),
\ee
which reproduces the known result \eqref{eq:FZZBC}. 
Note that $m-n + \frac{n}{q+1} = (m-n + \frac{m}{q})(\frac{q}{q+1})$, 
allowing us to see that $\langle{O}^{(q)}_{n,m}\rangle^{(1,1)}(s)$ from \eqref{eq:unitaryOnm}
and $\langle{O}^{(q-1)}_{m,n}\rangle^{(1,1)}(s)$ from \eqref{eq:unitaryOmn} are equal. 
This completes the demonstration of the equivalence between the $A_{q}$ and $A_{q-1}$ descriptions
for the one-point correlator on a disk.

The on-shell value of $\omega(\tilde x_\alpha)$ with 
$\tilde x_\alpha=\tilde \kappa \cosh(\pi \tilde b \tilde s_\alpha)$ reads
\begin{equation*}
\omega(\tilde x_\alpha)_* 
	= -\int_{0}^{\infty}dl\,e^{-l\tilde x_\alpha}\int_0^\infty dx \int_{i\mathbb{R}} dy \ e^{lQ_{q+1}(y,v(x))}
	\propto \tilde\kappa^{q/(q+1)}\cosh(\pi \tilde s_\alpha \tilde b^{-1}).
\end{equation*}
The one-point boundary correlation numbers $\ll B\rr^{(k,1)}$ given by \eqref{eq:BdyC:1k}
vanish for each $k\geq2$, since $\tilde s_\alpha=\tilde s+i\alpha \tilde b$.
As we have obtained \eqref{eq:BulkOnept:OmnL}, 
The bulk-one point correlation numbers with $(k,1)$ boundary
$\ll O_{n,m}\rr^{(k,1)}(s)$ reduce to a sum of $\ll O_{n,m}\rr^{(1,1)}(s_\alpha)$ as in \eqref{eq:BC:k1}. 
The bulk-boundary two-point correlation  \eqref{eq:BBC:k1} is
\bea
 \ll B O_{n,m} \rr^{(k,1)}(s) 
&\propto \sum_{\alpha=-(k-1):2}^{k-1} \xi_\alpha (\tilde x)
	\frac{\sinh\left(\pi \tilde b \tilde s_\alpha \frac{m(q+1)-nq}{q+1}\right)}{\sinh\left(\pi \tilde b \tilde s_\alpha\right)},	
\ena
with $P_{n,m}=\frac{\tilde b}{2}\frac{m(q+1)-nq}{q+1}= \frac{b}{2} \frac{m(q+1)-nq}{q}$ ($n\leq m$) and $B=B_{k,1}$,
which is proportional to $R(P_{n,m},k\tilde b,s)$. Note that $P_{n,m}$ is the same as 
$P_{m,n}$ in the original description $A_{q-1}$.

\section{Lee-Yang model: $\cM(2,5)$ on disk}

\subsection{Frobenius manifold $A_1$}

The Lee-Yang model is the simplest  for the Lee-Yang series $\cM(2,2q+1)$ on disk.
The one-dimensional coordinate  $v$ of the Frobenius manifold $A_1$ 
describes the Lee-Yang series \cite{BDM}.
We summarize the original Frobenius manifold description in this subsection.

The second order polynomial $Q_2= y^2+v$ is used to find 
the Douglas action for $\cM(2,5)$
\bea
	S_{2/5}= t^{1,1}\frac{v^2}{2}+t^{1,2}v-\frac{v^4}{4},
\ena 
where the KdV parameters have the  scaling dimensions $[t^{1,1}]=1$ and $[t^{1,2}]=3/2$.
There is no resonance term between the two couplings up to the order of our interest, 
so that the KdV parameter $t^{1,n}$ is identical to the Liouville coupling constant $\lambda^{1,n}$.
A derivative of the action $S_{2/5}$ with respect to $v$ gives
\bea \label{DefofX}
	\frac{\pd S_{2/5}}{\pd v}&= t^{1,2} - x,  \quad
	 \quad x= v^3 - t^{1,1}v
\ena
where $v$ is defined as the function of $x$. 
The  generating function $\mathcal{W}(l)$  is given in terms of integration over $x$
whose lower limit of integration range is identified as $\lambda^{1,2}$
\bea
	\mathcal{W}(l)&= \int_{\lambda^{1,2}}^\infty dx \int_{i\mathbb{R}} dy \ e^{ l ( y^2+v(x)) }.
\ena

The bulk one-point correlation numbers  on disk with $(1,1)$ boundary condition 
 are given as 
\bea \label{O1n:25}
	\ll O_{1,n} \rr^{(1,1)}_l
	&\propto \kappa^{(5-2n)/2}K_{(5-2n)/2}(\kappa l) \quad (n=1, 2)\,.
\ena
After Laplace transform we have the correlation numbers
reads
\be
	\ll O_{1,n}\rr^{(1,1)}(s) \propto \kappa^{(5-2n)/2} \cosh\left(\pi b s \frac{5-2n}{2}\right),
\ee
in agreement with \eqref{eq:FZZBC}. 
In addition, the resolvent $\omega(x_\pm)$ reads
\bea
	\omega(x_\pm)_*
	&=2\sqrt{2} \pi\ i\ \mu^{5/4}  \cosh\left(\frac{5}{2}\pi b s_\pm\right)\nonumber\\
	&\propto \kappa^{5/2} \cosh\left(\pi s_\pm/b\right)
\ena
where  $x_\pm=\kappa \cosh(\pi b s_\pm)$ and $s_{\pm}=s\pm i b$.
Therefore, the one-point boundary correlation number $\ll B\rr^{(1,2)}=0$ 
according to \eqref{eq:BdyC:1k}.
For  the $(1,2)$ boundary, one has the bulk one-point correlation numbers 
$\ll O_{1,n}\rr^{(1,2)}(s)
=\sum_{\pm}\ll O_{1,n}\rr^{(1,1)}(s_{\pm})$. 
Finally, the bulk-boundary two-point correlation  \eqref{eq:BBC:1k}
\bea
 \ll B O_{1,n} \rr^{(1,2)}(s) 
&\propto R(P_{1,n},2b,s) 
\ena
with $P_{1,n}=\frac{b}{2}\frac{5-2n}{2}$ shows that $B=B_{1,2}$.

\subsection{Dual Frobenius manifold $A_4$ and its problem}

The dual description is given by  the four-dimensional  Frobenius manifold $A_4$
with $Q_5= y^5+u_1y^3+u_2y^2+u_3y +u_4$.
The relation between $u_i$'s and the flat coordinates $v_i$'s is given by
\bea
	u_1=v_1,\quad u_2=v_2,\quad u_3=v_3+(v_1)^2/5,\quad u_4=v_4+v_1v_2/5.
\ena
The dual Douglas action reads
\bea
	S_{5/2}= t^{1,1}v_3+t^{2,1}v_1 - \frac{(v_1)^2v_3+v_1(v_2)^2 -10v_2v_4-5(v_3)^2}{10},
\ena
whose derivatives with respect to $v_i$ give
\be
\begin{aligned} \label{DefofX52}
	\frac{\pd S_{5/2}}{\pd v_1}&=t^{2,1} -x \quad \textrm{where} \quad x= \frac{2v_1v_3+(v_2)^2}{10},\\
	\frac{\pd S_{5/2}}{\pd v_2}&=v_4 -\frac{v_1v_2}{5}, \\
	\frac{\pd S_{5/2}}{\pd v_3}&= t^{1,1}+v_3 -\frac{(v_1)^2}{10},\\
	\frac{\pd S_{5/2}}{\pd v_4}&=v_2.
\end{aligned}
\ee
These equations imply that $v_2$ and $v_4$ should be irrelevant and thus they will be omitted hereafter.
Then $x$ is reduced to
\bea
	x= (v_1v_3)/5.	\label{ReDefofX}
\ena
The KdV parameters has no resonance up to the order of our interest, so that we have $t^{n,1}=\lambda^{n,1}$. 
We choose the on-shell value of $v_i$ as
\bea
	(v_1)^2_*=10\,\lambda^{1,1},\quad
	(v_3)_*=0.
\ena

The generating function for bulk correlations $\mathcal{W}(l)$ reads
\bea
	\mathcal{W}(l)&= \int_{\lambda^{2,1}}^\infty dx \int_{i\mathbb{R}} dy \ 
	e^{l(y^5+v_1y^3+(v_3+(v_1)^2/5)y)}.		\label{DefofW52}
\ena
A derivative of $\mathcal{W}(l)$ with respect to $\lambda^{2,1}$ yields
\bea \label{O21for52}
	\langle O_{2,1}\rangle^{(1,1)}_l
	&\propto \tilde{\kappa}^{1/5}K_{1/5}(\tilde{\kappa} l),
\ena
where $\tilde{\kappa}\propto\mu^{5/4}\propto\kappa^{5/2}$.
After Laplace tranformation, we can easily check
that $\langle O_{2,1}\rangle^{(1,1)}$ in $A_4$ description 
has the same value $\langle O_{1,2}\rangle^{(1,1)}$ in $A_1$ description.

However, a troublesome issue arises in other 
correlation numbers.
For example,  $\langle O_{1,1}\rangle^{(1,1)}_l$
is given as a derivative of $\mathcal{W}(l)$ with respect to $\lambda^{1,1}$.
\bea
	\langle O_{1,1}\rangle^{(1,1)}_l
	&=\int_{0}^\infty dx\int_{i\mathbb{R}} dy \ 
	l\frac{\pd Q_5}{\pd \lambda^{1,1}}  \
	e^{l(y^5+v_1y^3+(v_3+(v_1)^2/5)y)}\bigg|_*,
\ena
with
\be \label{t11derivs}
	\frac{\pd Q_5}{\pd \lambda^{1,1}}=\frac{5v_1y^3+(2(v_1)^2-5v_3)y}{5v_3+(v_1)^2}.
\ee
The relation among $v_1$, $v_3$ and $x$ is given by \eqref{ReDefofX}.
The result depends on  how to perform the $x$-integration.
If one performs $x$-integration in terms of $v_1$,  $dx=(v_3/5)dv_1$ 
while maintaining $v_3=0$, then the final result is proportional to $v_3$ and vanishes. 
If one fixes the $x$-integral as $dx=(v_1/5)dv_3$, 
the integral does neither vanish
nor agrees with the $b\leftrightarrow b^{-1}$ duality. 
The absence of the canonical description of $x$-integration 
is carried over into Lee-Yang series $\cM(2,2q+1)$. 
It should be noted that similar problem already appeared in non-unitary series \cite{Aleshkin:2017yty}. 
Therefore, one has to devise a proper way to perform $x$-integration 
to understand the duality for non-unitary series. 

Suppose we compute the resolvent
\bea
	\omega(x_\pm)_*
	&=- \int_0^\infty dl \ e^{-lx_\pm} \mathcal{W}(l)\bigg|_*.
\ena
To perform  $x$-integration to find $\mathcal{W}(l)$
in the dual picture one may perform $v_3$-integral from $0$ to $-\infty$,
regarding $v_1$ and $v_3$ as independent of each other
and get the following result.
\bea
	\omega(\tilde x_\pm)_*
	=&\frac{1}{5} \int_0^\infty dl \ e^{-l\tilde x_\pm}   \int_{i\mathbb{R}} dy \ \frac{v_1}{ly} \ 
	e^{l(y^5+v_1y^3+((v_1)^2/5)y)}\bigg|_*\\
	=& \int_0^\infty \frac{dl}{l} \ e^{-l\tilde x_\pm} \int_{i\mathbb{R}} dy \ y \ 
	e^{l(y^5+(v_1)_*y^3+((v_1)_*^2/5)y)}\nonumber\\
	& - \frac{\pd}{\pd v_1} \bigg[\int_0^\infty \frac{dl}{l^2} \int_{i\mathbb{R}}  \frac{dy}{y^2}    
	 \ e^{-l\tilde x_\pm+lQ_5}\bigg]_*\\
	\propto&\,\, \tilde{\kappa}^{2/5}\cosh(\pi s_\pm/\tilde{b})
	+\frac{\pd}{\pd v_1}\left[\dots\right]
\ena
where 
$x_\pm=\tilde{\kappa}\cosh (\pi \tilde{b} s_\pm)$.
The first term is what we want from the perspective of $b\leftrightarrow b^{-1}$ duality.
This shows that one needs to find a way to compensate the unpreferable second term 
by properly choosing the $x$-integration. 

\section{Conclusion}

The $b\leftrightarrow b^{-1}$  duality of the minimal Liouville gravity  $\cM(q,p)$ 
was examined in the matrix model approach 
using the original and its dual Frobenius manifolds $A_{q-1}$ and $A_{p-1}$.
We calculate the bulk one-point  and bulk-boundary two 
point correlation numbers on a disk 
and found that the duality holds for the unitary series $\cM(q,q+1)$.  
The $(1,k)$  boundary condition in the original picture 
corresponds to the $(k,1)$ boundary in the  dual picture.
The resonance transformation in the dual picture is also used.   

In contrast, for non-unitary series the duality is not easy to check. 
The correlation numbers without $x$-integration
show the duality. 
However, the correlations with the presence of the $x$-integration 
depends on the choice of the integration path.
This shows that one has to find a canonical prescription of contour integrations
in the non-unitary series. 

In the paper, we consider the boundary conditions of the type $(1,k)$ in the original 
and $(k,1)$ in the dual descriptions. 
One may wonder if one can  extend the results with mixed boundary conditions of the  type $(k,l)$
since according to the brane picture in \cite{SS} one has 
\be
	 | (k,l);s  \rangle  = \sum_{\alpha_1=-(k-1):2}^{k-1}   \sum_{\alpha_2=-(l-1):2}^{l-1}
	 | (1,1);  s_{\alpha_1,\alpha_2}  \rangle 
\ee
with  $ s_{\alpha_1,\alpha_2}= s +i\alpha_1  \tilde b + i\alpha_2 b$.
This is expected to appear after  an extension of the formalism
by combining both the original and dual Frobenius descriptions, which will be presented in the near future.\\

\noindent \textbf{Acknowledgements.} 
The work was partially supported by National Research Foundation of Korea grant 2017R1A2A2A05001164.

\appendix

\section{Frobenius structure}

We recall some facts and collecting some useful quantities about $A_{q-1}$ Frobenius manifold. 
The reader is referred to the appendix A.1. of \cite{BDM} for a detailed summary of these notions.

In the flat co-ordinates of $A_{q-1}$ Frobenius manifold, the metric takes the form
\begin{equation}
\eta_{ab}  = \delta_{a+b,q},
\end{equation}
where the metric is used to raise and lower these indices ($v^a = v_{q-a}$ for instance). 
The Frobenius algebra structure constants in these co-ordinates are given by
\begin{equation}
C_{abc} = -q \underset{y=\infty}{\mathrm{Res}}\frac{\frac{\pd Q(y)}{\pd v^a}\frac{\pd Q(y)}{\pd v^b}\frac{\pd Q(y)}{\pd v^c}}{Q'(y)}.
\end{equation}
The structure constants are symmetric under all permutation of indices and are defined in such a way that $C_{1ab} = \eta_{ab} = \delta_{a+b,q}$.

For our purposes it suffices to know the structure constants and various other quantities only along the line 
$v_{k}=0$ ($k>1$). All quantities evaluated on this line will be indicated by a subscript $\vec{v}$. 
The structure constants on this line are

\begin{equation*}
C_{abc}|_{\vec{v}} =\begin{cases}
\p{\frac{-v_1}{q}}^{\frac{a+b+c-q-1}{2}} &\text{if $\frac{a+b+c-q-1}{2} \in \{0\}\cup\mathbb{N}$, $1\leq a+b+c - 2\min(a,b,c)\leq q-1$,}\\
0 &\text{otherwise.}
\end{cases}
\end{equation*}
Particularly useful is
\begin{equation}\label{eq:strucconst}
C_{(q-1)ab} = \delta_{ab}\p{\frac{-v_1}{q}}^{a-1}.
\end{equation}
On the line $v_{k}=0$ ($k>1$), $Q$ and its derivative with respect to the flat co-ordinates are given by
\begin{equation}
\begin{aligned}
Q(y)|_{\vec{v}} 
&=2\p{-i}^q\p{\frac{v_1}{q}}^{q/2}T_q\p{\frac{i}{2}\sqrt{\frac{q}{v_1}}y}.\\
\p{\frac{\pd Q(y)}{\pd v_a}}_{\vec{v}} &= (-i)^{q-a-1}\p{\frac{v_1}{q}}^{\frac{q-a-1}{2}}T'_{q-a}\p{\frac{i}{2}\sqrt{\frac{q}{v_1}}y},
\end{aligned}
\end{equation}
where $T_n(z)$ is the Chebyshev polynomial.
Finally, we note the derivatives of $\theta_{i,k}$ respect to the flat co-ordinates on $\vec{v}$:
\begin{equation}\label{eq:dthetadv}
\p{\frac{\pd \theta_{i,k}}{\pd v_a}}_{\vec{v}} = \begin{cases}
\delta_{i,a}x_{i,k}\p{\frac{-v_1}{q}}^{\frac{kq}{2}} &\textrm{if $k$ is even,}\\
\delta_{i,q-a}y_{i,k}\p{\frac{-v_1}{q}}^{\frac{(k-1)q}{2} + i} &\textrm{if $k$ is odd,}
\end{cases}
\end{equation}
where
\begin{equation}
x_{i,k} = \frac{1}{\p{i/q}_{\frac{k}{2}}\p{\frac{k}{2}}!}, \hspace{0.5cm}y_{i,k} = \frac{-1}{\p{i/q}_{\frac{k+1}{2}}\p{\frac{k+1}{2}}!}.
\end{equation}

\section{Bessel function and Jacobi polynomial}

First we note the following facts about Chebyshev polynomials:
\begin{equation}\label{eq:ChebySinh}
T_m(i\sinh x) = \begin{cases}
i^m \sinh (mx) &\text{if $m$ is odd,}\\
i^m \cosh (mx) &\text{if $m$ is even.}
\end{cases}
\end{equation}
Consider the following integral
\begin{equation}
I = \int_{i\mathbb{R}} T'_m(y)e^{-zT_q(y)}dy.
\end{equation}
Let $q$ be even. If $m$ is even, then $T'_m(y)$ is an odd polynomial, making the integrand an odd function, thus causing the integral to vanish. If $m$ is odd, then the integrand is an even function, and after a substition $y = i\sinh x $ and using \eqref{eq:ChebySinh}, $I$ can be evaluated using the following integral representation of the modified Bessel function of the second kind:
\begin{equation}
K_\nu(z) = \int_{0}^\infty ds \cosh(\nu s)e^{-z \cosh s}.
\end{equation}
Now let $q$ be odd. We again substitute  $y = i\sinh x $. 
We can then use the integral representations
\begin{equation}
K_\nu(z) =
\begin{cases} 
	\displaystyle{\sec(\nu\pi/2)\int_{0}^\infty ds \cosh(\nu s)\cos(z\sinh s)}\quad\textrm{for }  {m} \textrm{ odd},\\ 
	\vspace{-0.7em}\\
	\displaystyle{\csc(\nu\pi/2)\int_{0}^\infty ds \sinh(\nu s)\sin(z\sinh s)},\quad\textrm{for }  {m} \textrm{ even}.
\end{cases}
\end{equation}

To perform the $x$-integration when one computes the correlation numbers, 
the following integration formula is essential:
\begin{equation}\label{eq:JacobiBesselInt}
l\int_{\mu}^{\infty}(\sqrt{x})^{-\nu}\mu^kP_k^{(0,-\nu)}\p{\tfrac{2x}{\mu}-1}K_\nu(\sqrt{x} l)dx = 2\sqrt{\mu}^{2k+1-\nu}K_{2k+1-\nu}(\sqrt{\mu} l).
\end{equation}
where $0<|\nu|<1$ for our purposes, and we recall that $K_\nu(x)= K_{-\nu}(x)$. To prove this formula, we first observe that the Jacobi polynomial appearing above can be written as a polynomial in $(x-\mu)$ as follows:
\begin{equation}\label{eq:JacobiExp}
\mu^kP_k^{(0,-\nu)}\p{\tfrac{2x}{\mu}-1}=\sum_{i=0}^{k}\binom{k}{i}\frac{1}{i!}(-\nu+k+1)_i\mu^{k-i}(x-\mu)^i.
\end{equation}
We now perform the $x$-integration term by term, using
\begin{equation}\label{eq:BesselInt}
l\int_{\mu}^{\infty} \sqrt{x}^{\nu}(x-\mu)^{m}K_\nu(l \sqrt{x})dx = 2^{m+1}\Gamma(m+1)l^{-m}\sqrt{\mu}^{m+\nu+1}K_{m+\nu+1}(l\sqrt{\mu}),
\end{equation}
which is proved by using the identity $\pd_xK_\nu(x)= -K_{\nu-1}(x)-(\nu/x)K_\nu(x)$ recursively. 
Finally, the linear combination of Bessel functions thus obtained can be summed using the following identity:
\begin{equation}\label{eq:BesselSum}
\sum_{i=0}^n 2^i\binom{n}{i}(\nu+n)_i\frac{K_{\nu+i}(x)}{x^i}=K_{\nu+2n}(x),
\end{equation}
which is proved by using the identity $K_{\nu}(x)+\frac{2(\nu+1)}{x}K_{\nu+1}(x)= K_{\nu+2}(x)$ recursively. 
To summarize the derivation of \eqref{eq:JacobiBesselInt}:
$$
\begin{aligned}
&l\int_{\mu}^{\infty}(\sqrt{x})^{-\nu}\mu^kP_k^{(0,-\nu)}\p{\tfrac{2x}{\mu}-1}K_\nu(\sqrt{x} l)dx\\
&=\sum_{i=0}^{k}\left[\binom{k}{i}\frac{1}{i!}(-\nu+k+1)_i\mu^{k-i}l\int_{\mu}^{\infty}(\sqrt{x})^{-\nu}(x-\mu)^iK_\nu(\sqrt{x} l)dx\right] &&\textrm{using \eqref{eq:JacobiExp}}\\
&= 2\sqrt{\mu}^{2k+1-\nu}\sum_{i=0}^{k}\left[2^i\binom{k}{i}(-\nu+k+1)_i\mu^{k-i}l^{-i}\frac{K_{i+1-\nu}(\sqrt{\mu} l)}{\p{\sqrt{\mu} l}^i}\right]&&\textrm{using \eqref{eq:BesselInt}}\\
&=2\sqrt{\mu}^{2k+1-\nu}K_{2k+1-\nu}(\sqrt{\mu} l)&&\textrm{using \eqref{eq:BesselSum}}.
\end{aligned}
$$

Finally, we note the following identity, which appears when we perform the Laplace transform ($l$-integration),
\be	\label{BesselCosh}
	\cosh (\nu t)
	= -\frac{\sin(\nu \pi)}{\pi} \int^\infty_0 \frac{dl}{l} \ e^{-\alpha l}  K_\nu(\beta l),
\ee
with $t = \acosh(\alpha/\beta)$. Differentiating with respect to $\alpha$ gives
\be
	\frac{\nu \sinh(\nu t)}{\beta \sinh (t)}
	= \frac{\sin(\nu \pi)}{\pi} \int^\infty_0 dl \ e^{-\alpha l}  K_\nu(\beta l).
\ee

\section{Orthogonality of two-point correlators on the sphere}

The first part of this section is largely a summary (with some notations modified for convenience) of section 3.2 of 
\cite{Belavin}, where the two-point correlator on a sphere was calculated, and the resonance transformations were found for unitary models $\cM(q,q+1)$. We refer the reader to that paper for further explanations. In the second part, we do a similar calculation for $\cM(q+1,q)$.

\subsection{$A_{q-1}$ description}

The two-point correlator on a sphere is given by
\begin{equation*}
\langle O_{m_1,n_1}O_{m_2,n_2}\rangle = \int_{0}^{v_*}dv_1 C_{a,b,q-1}\frac{\pd S^{m_1,n_1}}{\pd v_a}\frac{\pd S^{m_2,n_2}}{\pd v_b} = \sum_{a=1}^{q-1}(-q)^{1-a}\int_{0}^{v_*}dv_1v_1^{a-1}\frac{\pd S^{m_1,n_1}}{\pd v_a}\frac{\pd S^{m_2,n_2}}{\pd v_a}
\end{equation*}
where \eqref{eq:strucconst} has been used. Using \eqref{eq:dthetadv}, we find that\footnote{Since all quantities are calculated on the line $v_{k}=0$ ($k>1$), we will suppress the subscript indicating this.}
\begin{equation}\label{eq:dsleadingorder_1}
\frac{\pd S^{m,n}}{\pd v_a} \propto \begin{cases}
\delta_{m,a}\p{v_1^{\frac{m-n}{2}q}+\dots} &\textrm{if $(m-n)$ is even}\\
\delta_{m,q-a}\p{v_1^{\frac{m-n-1}{2}q + m}+\dots} &\textrm{if $(m-n)$ is odd}
\end{cases}
\end{equation}
where using $S_{q/(q+1)} = S_0 + \sum_{m,n}\lambda^{m,n}S^{m,n} + O(\lambda^2)$, we see that
the dotted terms contain positive integer exponents of the cosmological constant $\mu$.

There are three cases to be considered: $(m_1-n_1)$ and $(m_2-n_2)$ are both even, they are both odd and they have opposite parity. It can be easily checked by dimensional analysis that when  $(m_1-n_1)$ and $(m_2-n_2)$ have opposite parity the two-point correlator is analytic (i.e., a non-negative integer power) in $\mu$, and must thus be discarded as non-universal \cite{BDM}. The only non-trivial constraints then arise from demanding the orthogonality of the two-point correlator in the cases when $(m_1-n_1)$ and $(m_2-n_2)$ are either both even, or both odd.

First consider the case when $m_1-n_1$ and $m_2-n_2$ are even. Denote $\frac{\pd S^{m,n}}{\pd v_a} = \delta_{m,a}f^{m,n}$. Then
\begin{equation*}
\langle O_{m_1,n_1}O_{m_2,n_2}\rangle \propto \delta_{m_1,m_2} \int_{0}^{v_*}v_1^{\frac{m_1+m_2}{2}-1} f^{m_1,n_1}f^{m_2,n_2}dv_1.
\end{equation*}
We already see that the two-point correlator vanishes if $m_1\neq m_2$, but we still need to explicitly require that it vanishes if $n_1\neq n_2$. Define $t = 2\p{\frac{v_1}{v_*}}^q - 1 \equiv  \frac{2x}{\mu} - 1$. With this redefinition
\begin{equation*}
\langle O_{m_1,n_1}O_{m_2,n_2}\rangle \propto \delta_{m_1,m_2} v_*^{m_1}\int_{-1}^{1}(1+t)^{\frac{m_1+m_2}{2q}-1} f^{m_1,n_1}f^{m_2,n_2}dt.
\end{equation*}
Orthogonality now forces the $f^{m,n}$'s to be proportional to a particular class of Jacobi polynomials: \begin{equation*}
f^{m,n} \propto P^{(0,\frac{m}{q}-1)}_{k(m,n)}(t),
\end{equation*}
where $k(m,n)$ is some integer. Knowing that $P^{(0,b)}_{k}(t) \propto t^k +(\textrm{lower-order terms in $t$})$, and comparing with \eqref{eq:dsleadingorder_1}, we see that $k(m,n) = \frac{m-n}{2}$. Putting all this together, we find that
\begin{equation} \label{eq:C2}
\frac{\pd S^{m,n}}{\pd v_a} \propto \delta_{m,a} {\mu}^\frac{m-n}{2}P_{\frac{m-n}{2}}^{(0,\frac{m}{q}-1)}\p{\tfrac{2x}{\mu}-1}
\quad \textrm{when $(m-n)$ is even.}
\end{equation}
The exponent of $\mu$ can be verified by dimensional analysis.

Now let both $(m_1-n_1)$ and $(m_2-n_2)$ be odd. 
Denote $\frac{\pd S^{m,n}}{\pd v_a} = \delta_{m,q-a}g^{m,n}$. Then
\begin{equation*}
\begin{aligned}
\langle O_{m_1,n_1}O_{m_2,n_2}\rangle &\propto\delta_{m_1,m_2}  \int_{0}^{v_*}v_1^{q-1-\frac{m_1+m_2}{2}} g^{m_1,n_1}g^{m_2,n_2} dv_1 \\ 
&\propto \delta_{m_1,m_2} v_*^{q-m_1}  \int_{-1}^{1}\p{1+t}^{-\frac{m_1+m_2}{2q}} g^{m_1,n_1}g^{m_2,n_2} dt.
\end{aligned}
\end{equation*}
Call $g^{m,n} = \p{1+t}^{\frac{m}{q}}h^{m,n}$. Then
\begin{equation*}
\langle O_{m_1,n_1}O_{m_2,n_2}\rangle \propto \delta_{m_1,m_2} v_*^{q-m_1}  \int_{-1}^{1}\p{1+t}^{\frac{m_1+m_2}{2q}}h^{m_1,n_1}h^{m_2,n_2} dt.
\end{equation*}
Following a similar line of reasoning as in the even case, we conclude that $h^{m,n} \propto P_{\frac{m-n-1}{2}}^{(0,\frac{m}{q})}(t)$. Putting it all together, we find that
\begin{equation} \label{eq:C3}
\frac{\pd S^{m,n}}{\pd v_a} \propto \delta_{m,q-a}x^\frac{m}{q} {\mu}^\frac{m-n-1}{2}P_{\frac{m-n-1}{2}}^{(0,\frac{m}{q})}\p{\tfrac{2x}{\mu}-1}
\quad \textrm{when $(m-n)$ is odd.}
\end{equation}

\subsection{$A_{q}$ description}

Consider again the two-point correlator on the sphere:
\begin{equation*}\begin{aligned}
\langle O_{n_1,m_1}O_{n_2,m_2}\rangle &= \int_{0}^{v_*}dv_1 C_{a,b,q}\frac{\pd S^{n_1,m_1}}{\pd v_a}\frac{\pd S^{n_2,m_2}}{\pd v_b}\\ &= \sum_{a=1}^{q}(-q-1)^{1-a}\int_{0}^{v_*}v_1^{a-1}\frac{\pd S^{n_1,m_1}}{\pd v_a}\frac{\pd S^{n_2,m_2}}{\pd v_a}dv_1
\end{aligned}\end{equation*}
\begin{equation}\label{eq:dsleadingorder_2}
\frac{\pd S^{n,m}}{\pd v_a} \propto \begin{cases}
\delta_{n,a}\p{v_1^{\frac{m-n}{2}(q+1)}+\dots} &\textrm{if $(m-n)$ is even}\\
\delta_{n,q+1-a}\p{v_1^{\frac{m-n-1}{2}(q+1) + m}+\dots} &\textrm{if $(m-n)$ is odd}
\end{cases}
\end{equation}
where the dots now denote terms containing positive integer exponents of the dual cosmological constant $\tilde{\mu}$.

As before there are three cases depending on the parities of $(m_1-n_1)$ and $(m_2-n_2)$. The scaling dimension of the two-point correlator is
\begin{equation*}
\begin{aligned}
[\langle O_{n_1,m_1}O_{n_2,m_2}\rangle] &= \frac{2q+1}{q+1}-\p{\frac{2q+1-m_1(q+1)+n_1}{2(q+1)}}-\p{\frac{2q+1-m_2(q+1)+n_2q}{2(q+1)}}\\
&=\frac{m_1-n_1}{2}+\frac{m_2-n_2}{2}+\frac{n_1+n_2}{2(q+1)}.
\end{aligned}\end{equation*}
When $(m_1-n_1)$ and $(m_2-n_2)$ have opposite parities, we see that $n_1+n_2 = q+1$, and that the the scaling dimension above is an integer. Once again, when the result of an expectation value turns out to be analytic in $\tilde\mu$, the result must be discarded as being non-universal. This leaves us with cases where $(m_1-n_1)$ and $(m_2-n_2)$ have the same parities.

Consider the case when $(m_1-n_1)$ and $(m_2-n_2)$ are even. Denote $\frac{\pd S^{n,m}}{\pd v_a} = \delta_{n,a}\tilde f^{n,m}$. Then
\begin{equation*}
\langle O_{n_1,m_1}O_{n_2,m_2}\rangle \propto \delta_{n_1,n_2} \int_{0}^{v_*}v_1^{\frac{n_1+n_2}{2}-1} \tilde f^{n_1,m_1}\tilde f^{n_2,m_2}dv_1.
\end{equation*}
It helps to define $\tilde t = 2\p{\frac{v_1}{v_*}}^{q+1} - 1 \equiv  \frac{2\tilde x}{\tilde\mu} - 1$. In terms of this new variable,
\begin{equation*}
\langle O_{n_1,m_1}O_{n_2,m_2}\rangle \propto \delta_{n_1,n_2} v_*^{n_1}\int_{-1}^{1}(1+\tilde t)^{\frac{n_1+n_2}{2(q+1)}-1} \tilde f^{n_1,m_1}\tilde f^{n_2,m_2}d\tilde t.
\end{equation*}
We again see the emergence of the same family of Jacobi polynomials as in the previous section. Comparing with \eqref{eq:dsleadingorder_2}, we find that
\begin{equation}
\frac{\pd S^{n,m}}{\pd v_a}\propto
\delta_{n,a} \tilde{\mu}^\frac{m-n}{2}P_{\frac{m-n}{2}}^{(0,\frac{n}{q+1}-1)}\p{\tfrac{2\tilde{x}}{\tilde{\mu}}-1} 
\quad \textrm{when $(m-n)$ is even.}
\end{equation}

Finally, let $(m_1-n_1)$ and $(m_2-n_2)$ be both odd. Denote $\frac{\pd S^{n,m}}{\pd v_a}= \delta_{m,q+1-a}\tilde g^{n,m}$. Then
\begin{equation*}
\begin{aligned}
\langle O_{n_1,m_1}O_{n_2,m_2}\rangle &\propto\delta_{n_1,n_2}  \int_{0}^{v_*}v_1^{q-\frac{n_1+n_2}{2}} \tilde g^{n_1,m_1}\tilde g^{n_2,m_2} dv_1 \\ 
&\propto \delta_{n_1,n_2} v_*^{q-m_1+1}  \int_{-1}^{1}\p{1+\tilde t}^{-\frac{n_1+n_2}{2(q+1)}} \tilde g^{n_1,m_1}\tilde g^{n_2,m_2} d\tilde t.
\end{aligned}
\end{equation*}
Call $\tilde g^{n,m} = \p{1+\tilde t}^{\frac{n}{q+1}}\tilde h^{n,m}$. Then
\begin{equation*}
\langle O_{n_1,m_1}O_{n_2,m_2}\rangle \propto \delta_{n_1,n_2} v_*^{q-m_1+1}  \int_{-1}^{1}\p{1+\tilde t}^{\frac{n_1+n_2}{2(q+1)}} \tilde h^{n_1,m_1}\tilde h^{n_2,m_2} d\tilde t.
\end{equation*}
so that orthogonality implies $\tilde h^{n,m} \propto P_{\frac{m-n-1}{2}}^{(0,\frac{n}{q+1})}\p{\tilde t}$, and the exponent of $\tilde\mu$ can be fixed by matching dimensions. Finally, we get
\begin{equation}
\frac{\pd S^{n,m}}{\pd v_a}\propto\delta_{n,q+1-a} \tilde{x}^\frac{n}{q+1} \tilde{\mu}^\frac{m-n-1}{2}P_{\frac{m-n-1}{2}}^{(0,\frac{n}{q+1})}\p{\tfrac{2\tilde{x}}{\tilde{\mu}}-1} 
\quad \textrm{when $(m-n)$ is odd.}
\end{equation}


\end{document}